\documentclass[10pt]{article}

\usepackage{authblk}
\usepackage{setspace}

\usepackage{amsmath}
\usepackage{graphicx}
\usepackage{amssymb}
\usepackage{color,soul}
\usepackage[T1]{fontenc}
\usepackage{bm}
\usepackage{ae,aecompl}
\usepackage{nameref}

\providecommand{\keywords}[1]{\textbf{\textit{Keywords---}} #1}

\begin{document}



\title{An experimental proof that resistance-switching memory cells are not memristors}

\author[1]{J.~Kim}

\author[1]{Y. V. Pershin\footnote{Corresponding author, Email:pershin@physics.sc.edu}}

\author[2]{M.~Yin}

\author[1]{T.~Datta}

\author[3]{M. Di Ventra}

\affil[1]{Department of Physics and Astronomy, University of South Carolina, Columbia, South Carolina 29208, USA}
\affil[2]{Physics/Engineering Department, Benedict College, Columbia, South Carolina 29204, USA}
\affil[3]{Department of Physics, University of California, San Diego, La Jolla, CA 92093, USA}

\maketitle


\begin{abstract}
It has been suggested  that {\it all} resistive-switching memory cells are memristors. The latter  are hypothetical, ideal devices whose resistance, as originally formulated,
depends {\it only} on the net charge that traverses them.
Recently, an  {\it unambiguous} test has been proposed [J. Phys. D: Appl. Phys. {\bf 52}, 01LT01 (2019)] to  determine whether a given physical system is indeed a memristor or not.
Here, we experimentally apply such a test to both in-house fabricated Cu-SiO$_2$ and commercially available electrochemical metallization cells. Our results unambiguously show that electrochemical metallization memory cells are {\it not} memristors. Since the particular resistance-switching memories employed in our study share similar features with many other memory cells, our findings refute the claim that all resistance-switching memories are memristors.  They also cast doubts on the existence of ideal memristors as actual physical devices that can be fabricated experimentally. Our results then lead us to formulate two {\it memristor impossibility conjectures} regarding the impossibility of building a model of physical resistance-switching memories based on the memristor model.
\end{abstract}




\section{Introduction}

Although some publications~\cite{strukov08a,Wang19a} have claimed that the  memristor~\cite{chua71a} (in the ideal sense) has been found and
{\it all} resistance-switching memories are memristors~\cite{chua2011a}, several researchers have raised serious doubts about such claims~\cite{mouttet2012memresistors,meuffels2012fundamental,di2013physical,vongehr2015missing,sundqvist2017memristor,comment_19a}.
Indeed, the property of pinched hysteresis loops~\footnote{It is interesting to note that the hysteresis loops of resistance-switching memories are typically twisted, not pinched~\cite{comment_19a}.} alone (``If it's pinched it's a memristor\,''~\cite{chua2014if}) {\it cannot} serve as a good indicator of memristors since that property is shared by different types of experimentally-realizable devices (such as  memristive devices and systems whose memory depends on some internal degrees of freedom, other than the charge~\cite{chua76a}).

Remarkably, the most important characteristic of {\it any} memristor~\cite{chua71a}, namely, the functional dependence of its memory resistance (memristance), $R_M$, on {\it only} the net charge, $q$, that traverses it, $R_M(q)$, has {\it never been demonstrated experimentally}. However,
it is obvious that any claim of the ``memristor discovery'' must be based on the {\it experimental} measurement of $R_M(q)$, and not
merely on non-exclusive characteristics. Of course, since physical devices are not ideal, like any other circuit component (such as resistors, capacitors and inductors), the memristor (if found) would show some deviations from the ideal behavior, depending on the operation conditions. However, such deviations should be small, say, within 10~\% of the ideal $R_M(q)$ curve, for such a device to be a good representation of its ideal counterpart. Otherwise, it would be a totally different device altogether.




 In Ref.~\cite{pershin18a}, two of us (YVP and MD) have introduced a simple test to experimentally determine whether a resistor with memory is an ideal memristor
or something else. The main idea of the test is based on the duality property of a capacitor-memristor circuit whereby, for any initial resistance state of the memristor and any form and amplitude of the applied voltage, the final state of an ideal memristor must be identical to its initial state, if the capacitor charge finally returns to its initial value~\cite{pershin18a}. In other words, our test verifies the $R_M(q)$ dependence.

To prove the ideality, a scan over a large range of parameters (initial state, shape/magnitude of the applied voltage, etc.) is clearly required. To prove the opposite, however, even one or a few measurements demonstrating the absence of such a duality within the operating range of the device would be enough. Here, we report the results of more than 60 memristor tests performed over several kinds of resistance switching devices using triangular and rectangular voltage pulses of positive and negative polarities. The results of these tests are in mutual agreement, and in agreement with the theoretical modeling of Ref.~\cite{pershin18a}.

In the present paper, we experimentally apply the ideal memristor test~\cite{pershin18a} to in-house fabricated Cu-SiO$_2$ and commercially available electrochemical metallization cells as model resistance-switching devices. Electrochemical metallization cells (ECMs)~\cite{valov2011electrochemical} constitute a large family of resistance-switching devices based on the cation diffusion through a solid electrolyte. Typically, such cells exhibit bipolar resistance switching with thresholds. This property allows us to use the tested devices as representatives of the entire class of bipolar threshold-type resistance-switching cells~\cite{pershin11a}, which also includes valence change memory cells (VCMs)~\cite{waser2009redox}. Moreover, to demonstrate explicitly that VCMs are not memristors, we plotted the $R_M(q)$ dependence of TaO devices for several driving conditions
using an accurate model developed by the HP group~\cite{Strachan13a}. We have found that the device resistance can not even be  {\it approximately} described by the ideal relation $R_M =R_M(q)$.

Our  results show that the resistance-switching memories are {\it not}  memristors, and cast further doubts on the existence of ideal memristors as actual physical devices that can be fabricated in the laboratory or found in Nature. This leads us to formulate two {\it memristor impossibility conjectures}, namely that {\it i}) it is impossible
to accurately model physical resistance-switching memories by adding small corrections to the ideal memristor model, and {\it ii}) it is impossible to build a circuit combining {\it ideal} memristors with any other ideal two-terminal devices (resistors, capacitors, and inductors) that emulates realistically the response of experimentally-realizable resistance-switching memories.
\begin{figure}[b]
	\centering{(a)\includegraphics[width=35mm]{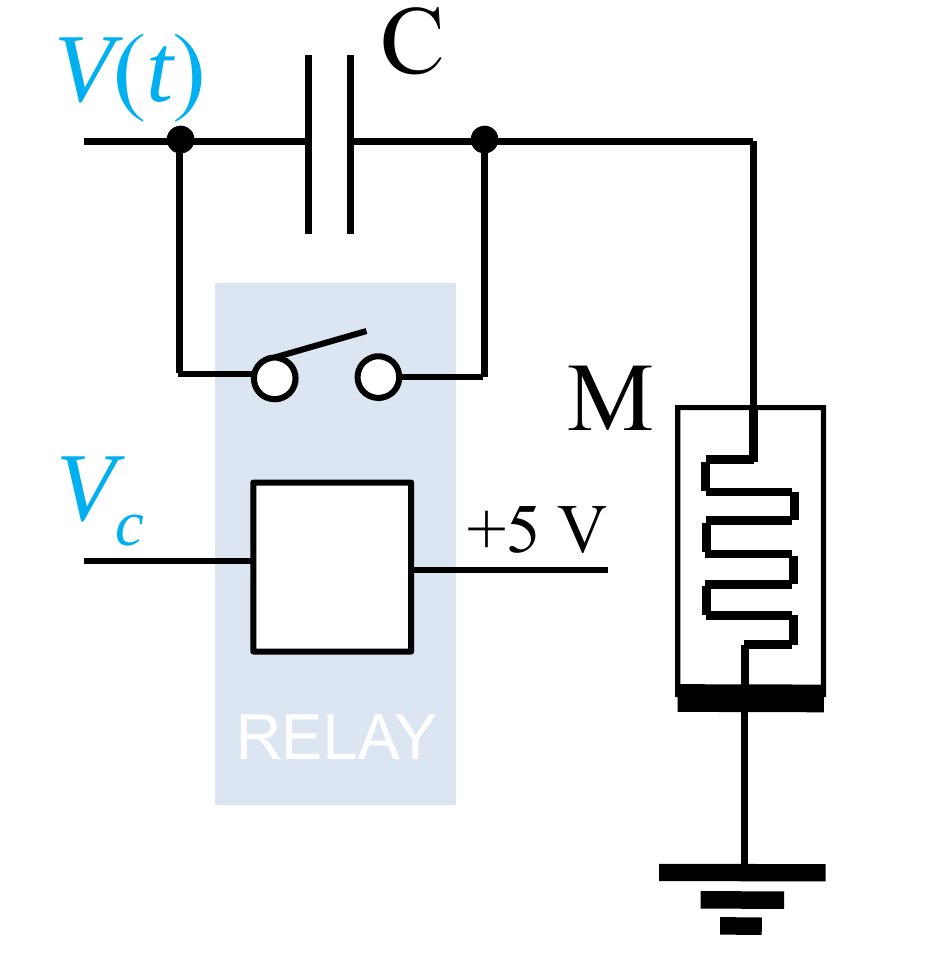}\;\;\;\;\;\;(b)\includegraphics[width=40mm]{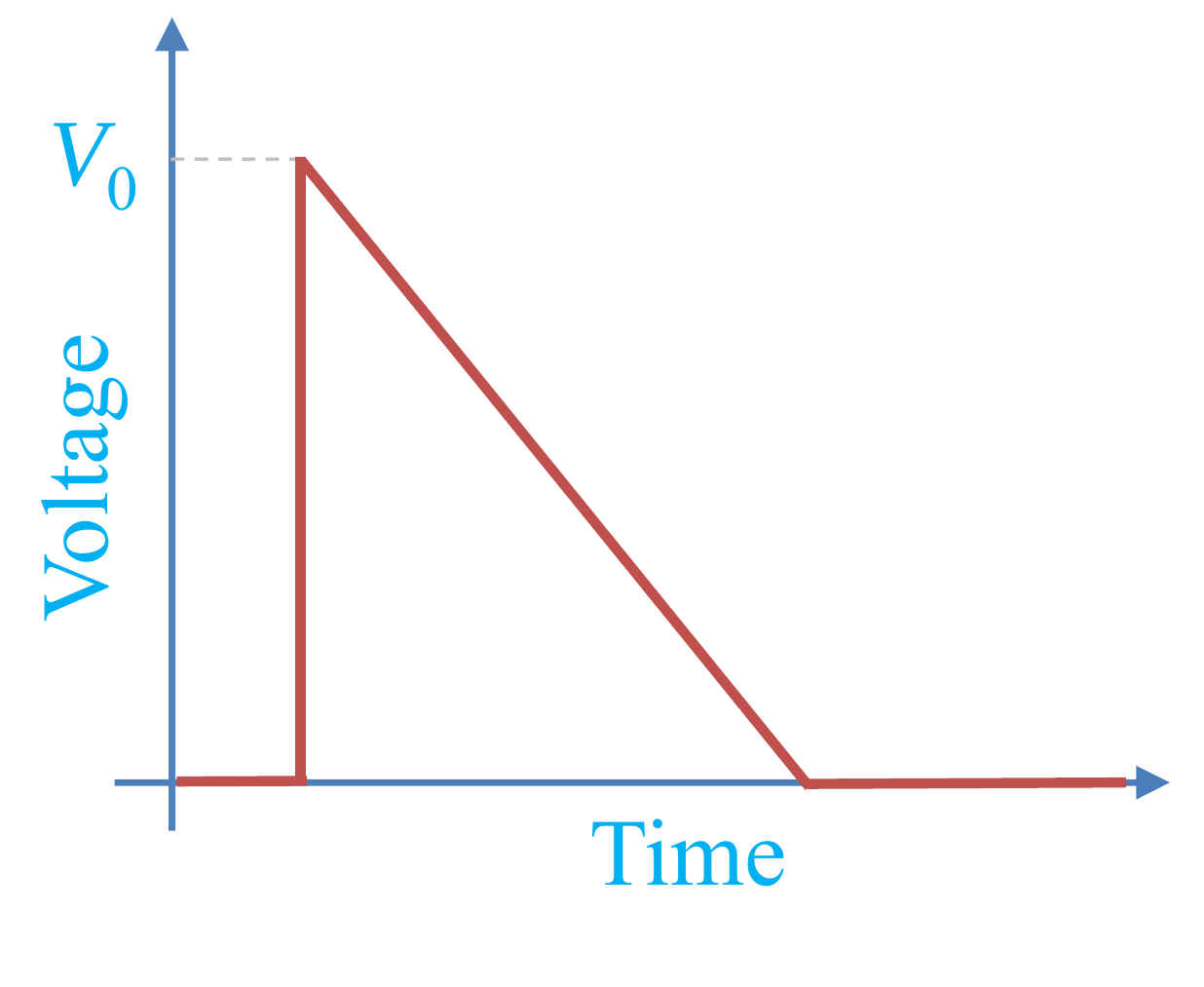}}
	\caption{(a) Capacitor-resistive memory circuit employed in our experiments. Here, a relay  is used to shunt the capacitor, C, to initialize and read the state of the resistive memory, M. (b) Shape of the testing voltage $V(t)$ employed in the present work.}
	\label{fig:1}
\end{figure}

\section{Methods} \label{sec:2}

Figure~\ref{fig:1}(a) shows the experimental circuit used to implement the ideal memristor test. We used a source measure unit (Keysight B2911A) to generate the test voltage signal $V(t)$ and control signal $V_c$. The unit is controlled by a code written in C Sharp. To initialize and measure the device state, we closed the relay (part number HI05-1A
66, Standex-Meder Electronics), thus connecting the tested device to the source measure unit directly. To run the test, we opened the relay and applied the test voltage signal (such as the triangular pulse in Figure~\ref{fig:1}(b)) across the capacitor (non-polarized 1~$\mu$F or 10~$\mu$F capacitor) connected in series with the tested device. In this work, the test was applied to $i$) in-house fabricated  Cu-SiO$_2$-based electrochemical metallization memory cells, and $ii$) commercially available ECMs by Knowm Inc. (BS-AF-W and M+SDC~Cr devices)~\cite{knowm}. Moreover, we used a precise model of TaO VCMs (another wide class of memory devices) to show that their characteristics are in striking disagreement with the memristor model.

\begin{figure}[t]
    \centering{\includegraphics[width=39mm]{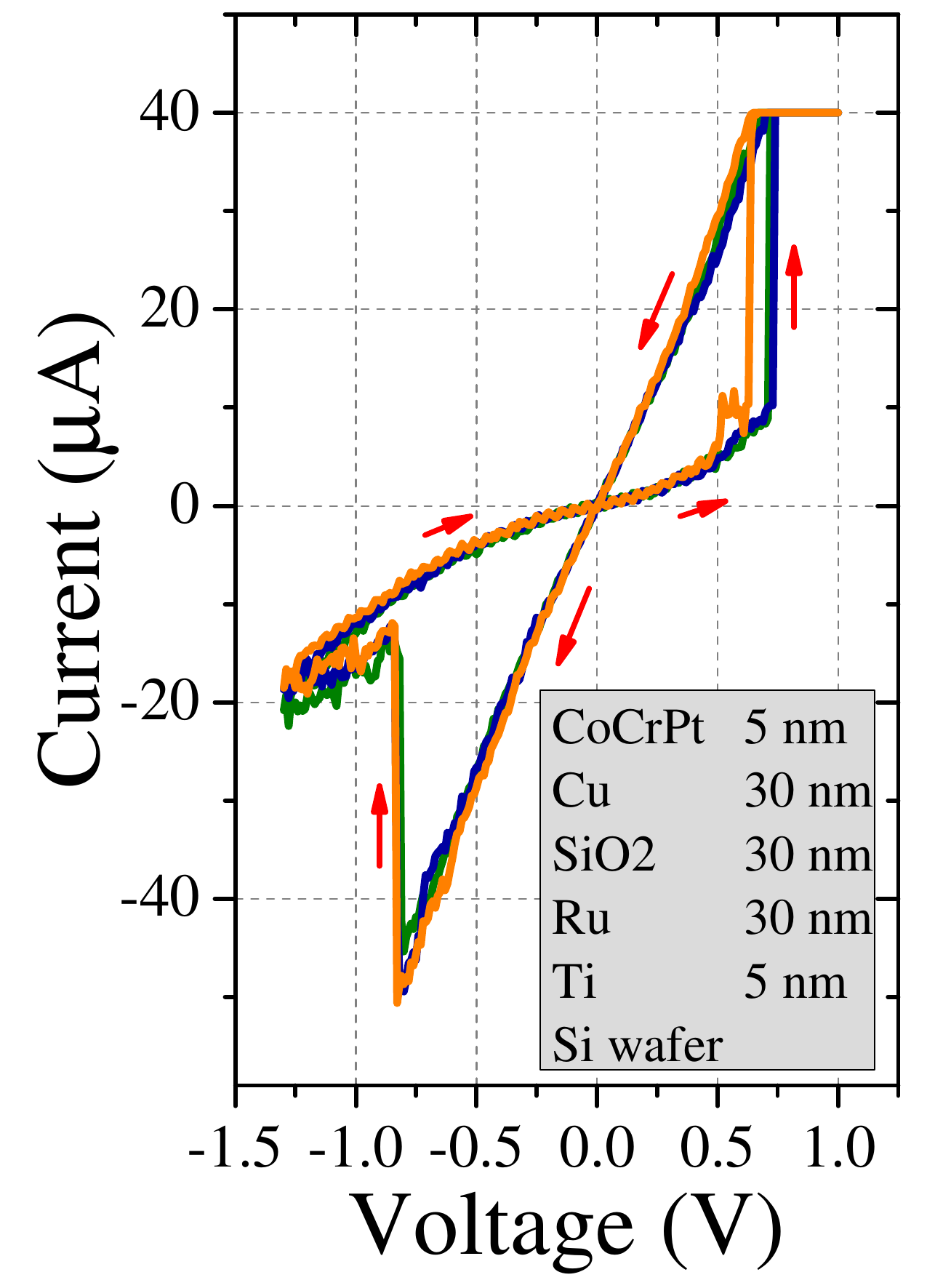} \includegraphics[width=39mm]{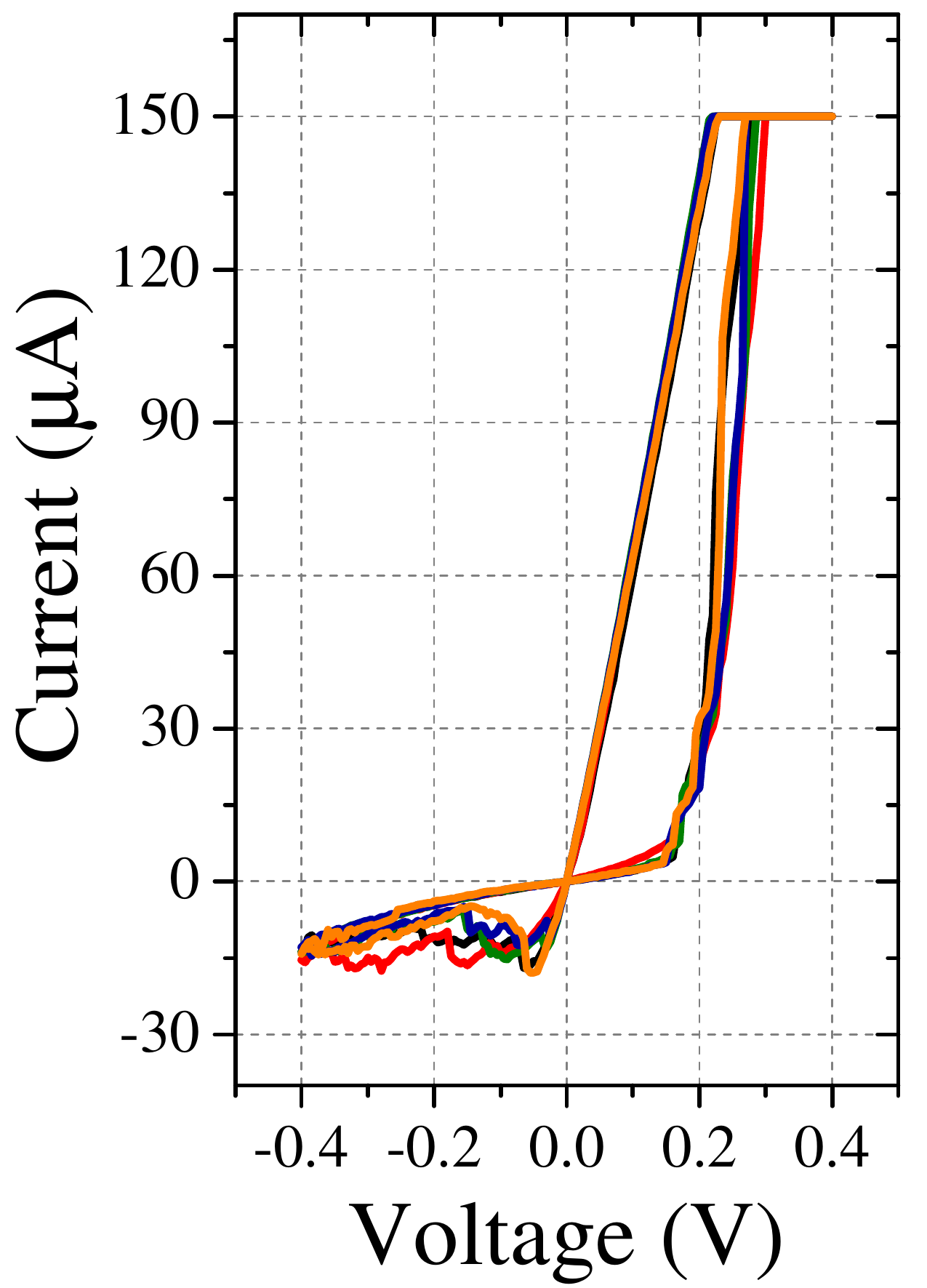}  \includegraphics[width=39mm]{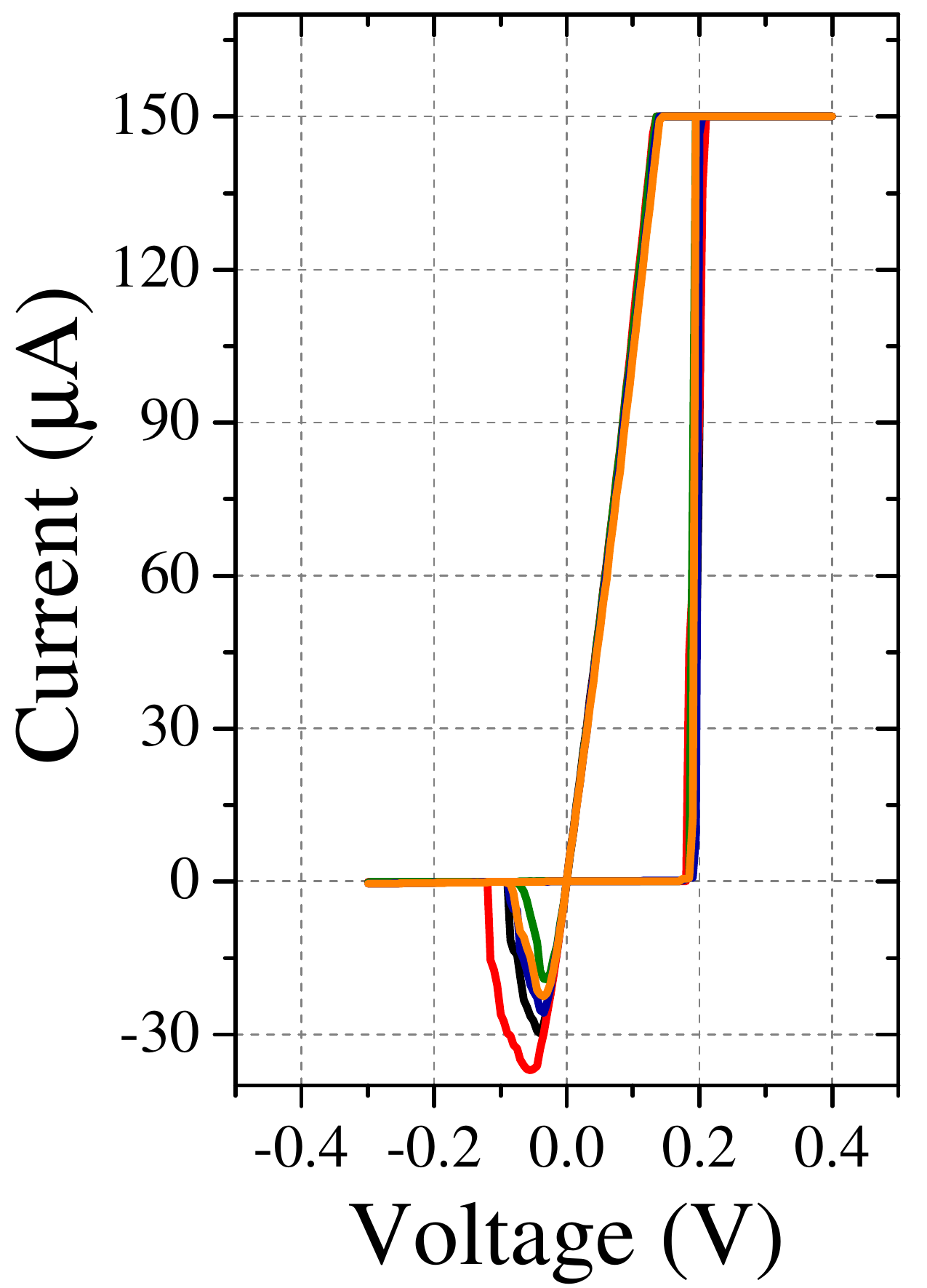}}
    (a) \hspace{35mm} (b) \hspace{30mm} (c)
	\caption{(a) Current-voltage characteristics of a Cu-SiO$_2$ device obtained using sweeps of 0 to +1 to -1.3 to 0~V with a positive current compliance of $40$~$\mu$A, and a negative current compliance of 1~mA. Inset: Schematic diagram of the Cu-SiO$_2$ device structure. (b) and (c):  Current-voltage characteristics of Knowm BS-AF-W (b) and M+SDC~Cr (c) devices obtained using $150$~$\mu$A current compliance.}
	\label{fig:2}
\end{figure}

The Cu-SiO$_2$ resistance-switching devices studied in this work were fabricated by sputtering deposition technique on the surface of a silicon wafer (substrate).
A thin adhesion layer ($5$~nm Ti)  was first formed on the surface of the substrate.
We used 30~nm Ru as an inert bottom electrode common for all devices. A 30-nm-thick SiO$_2$ layer was deposited using a shadow mask with $10\times 10$~mm square openings. The top Cu electrodes of 30~nm thickness were deposited on top of SiO$_2$  using another shadow mask with square and circular openings of various sizes (in this paper we present data for a device with a circular top electrode of $r=710$~$\mu$m). A 5~nm CoCrPt was used as a protective layer for the top electrodes (see inset of Figure~\ref{fig:2} for a schematic of the structure of memory cells). In order to dope SiO$_2$ with Cu atoms~\cite{schindler2007bipolar} the devices were subjected to 580~$^\circ$C, 1~hour annealing in He environment. After that, the samples were slowly cooled down to the room temperature. The result is a typical
resistance-switching memory cell with characteristics similar to many other experimental memory devices~\cite{pershin11a}. The ideal memristor test was implemented on several randomly selected devices showing stable switching behavior.

Our modeling results were obtained using an accurate model of TaO VCMs~\cite{Strachan13a} consisting of two equations
\begin{eqnarray}
  I &=& G(x,V_M)V_M \label{eq:1} \;\; , \\
  \dot{x} &=& A \sinh\left(\frac{V_M}{\sigma_{off}}\right)\exp\left( -\frac{x_{off}^2}{x^2}\right)\exp\left( \frac{1}{1+\beta I\;V_M}\right)H(-V_M) \nonumber \\
          &+& B \sinh\left(\frac{V_M}{\sigma_{on}}\right)\exp\left( -\frac{x^2}{x^2_{on}}\right)\exp\left( \frac{I\;V_M}{\sigma_{p}}\right)H(V_M) \;\; , \label{eq:2}
\end{eqnarray}
where Eq.~(\ref{eq:1}) is a generalized Ohms law, while Eq.~(\ref{eq:2}) describes the internal state dynamics. Here,
the state variable $x$ is the volume fraction of the oxygen-depleted channel with metallic transport, while the remaining fractional volume $1-x$ is insulating (with nonlinear transport)~\cite{Strachan13a}. According to Eq.~(\ref{eq:2}), the metallic channel either expands or shrinks depending on the bias polarity.  In Eq.~(\ref{eq:1}), the memductance $G(x,V_M)\equiv R_M^{-1}$ is
\begin{equation}\label{eq:3}
  G(x,V_M)=G_Mx +a \exp\left(b\sqrt{|V_M|}\right)(1-x),
\end{equation}
$A$, $B$, $\sigma_{off(on,p)}$, $x_{off(on)}$, $\beta$, $G_M$, $a$, $b$ are constants, and $H(...)$ is the Heaviside step function.
All simulations reported below were performed using the following set of parameter values~\cite{Ascoli17a}:
$A=10^{-10}$~s$^{-1}$, $B=10^{-4}$~s$^{-1}$, $\sigma_{off}=0.013$~V, $\sigma_{on}=0.45$~V, $\sigma_{p}=4\times 10^{-5}$~A$\;$V, $x_{off}=0.4$, $x_{on}=0.06$, $\beta=500$~A$^{-1}$V$^{-1}$, $G_M=0.025$~S, $a=7.2$~$\mu$S, $b=4.7$~V$^{-1/2}$.

\section{Results} \label{sec:3}

\subsection{Electrochemical metallization cells}

\subsubsection{Cu-SiO$_2$ devices}

Figure $\ref{fig:2}$(a) shows typical current-voltage characteristics of a selected Cu-SiO$_2$ device. This plot demonstrates a bipolar switching with well defined thresholds, and a hysteresis loop twisted at the origin. From this plot we estimate the following  parameters of our device:  $R_{on}\simeq 19.5$~k$\Omega$, $R_{off}\simeq 150$~k$\Omega$, $V_{t,+}\simeq 0.7$~V, and $V_{t,-}\simeq -0.8$~V. Here, $R_{on/off}$ are the boundary resistance values and $V_{t,+/-}$ are the threshold voltages.

\begin{figure*}[tb]
	\centering{\includegraphics[width=28mm]{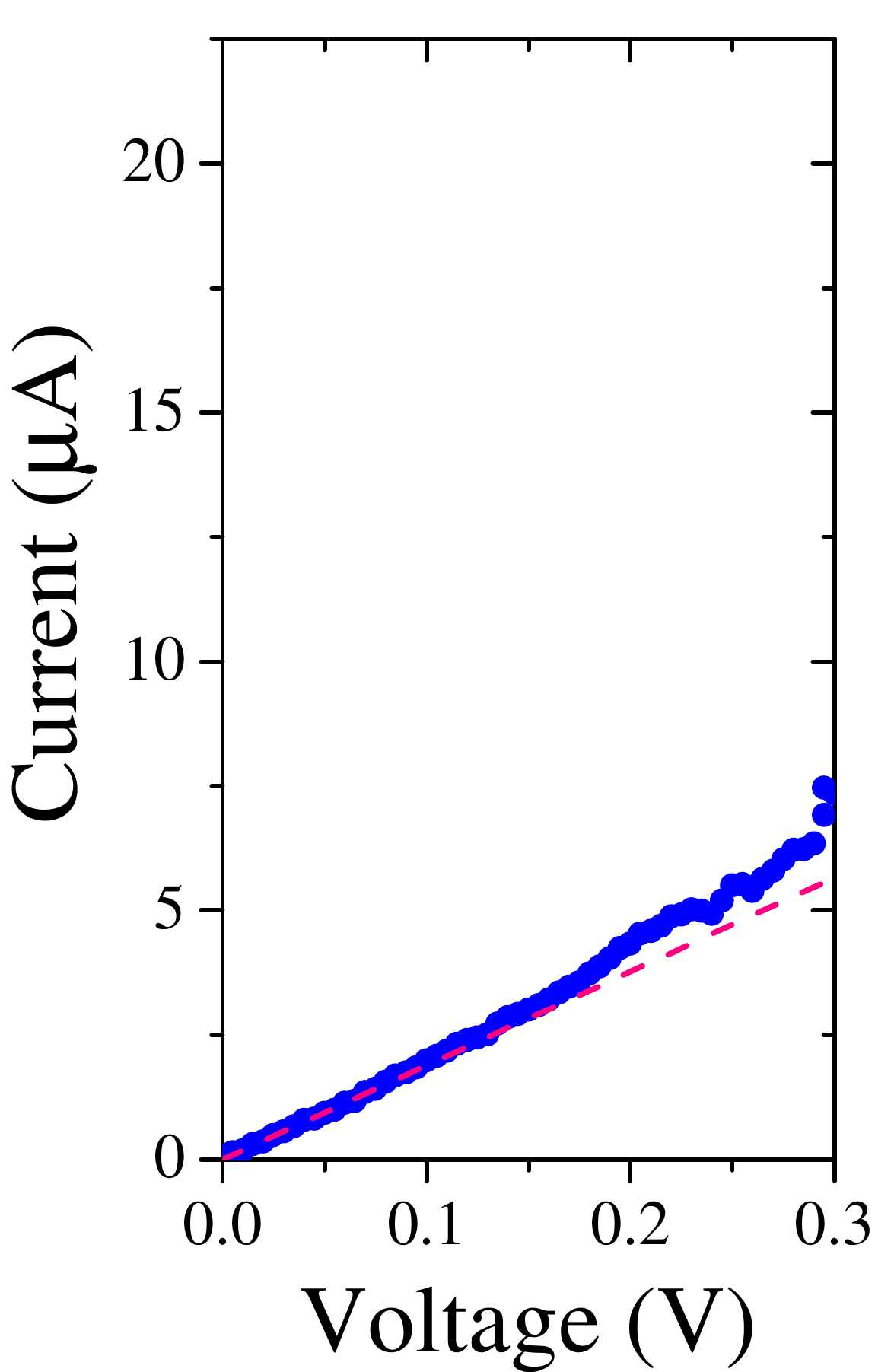} \; \includegraphics[width=58mm]{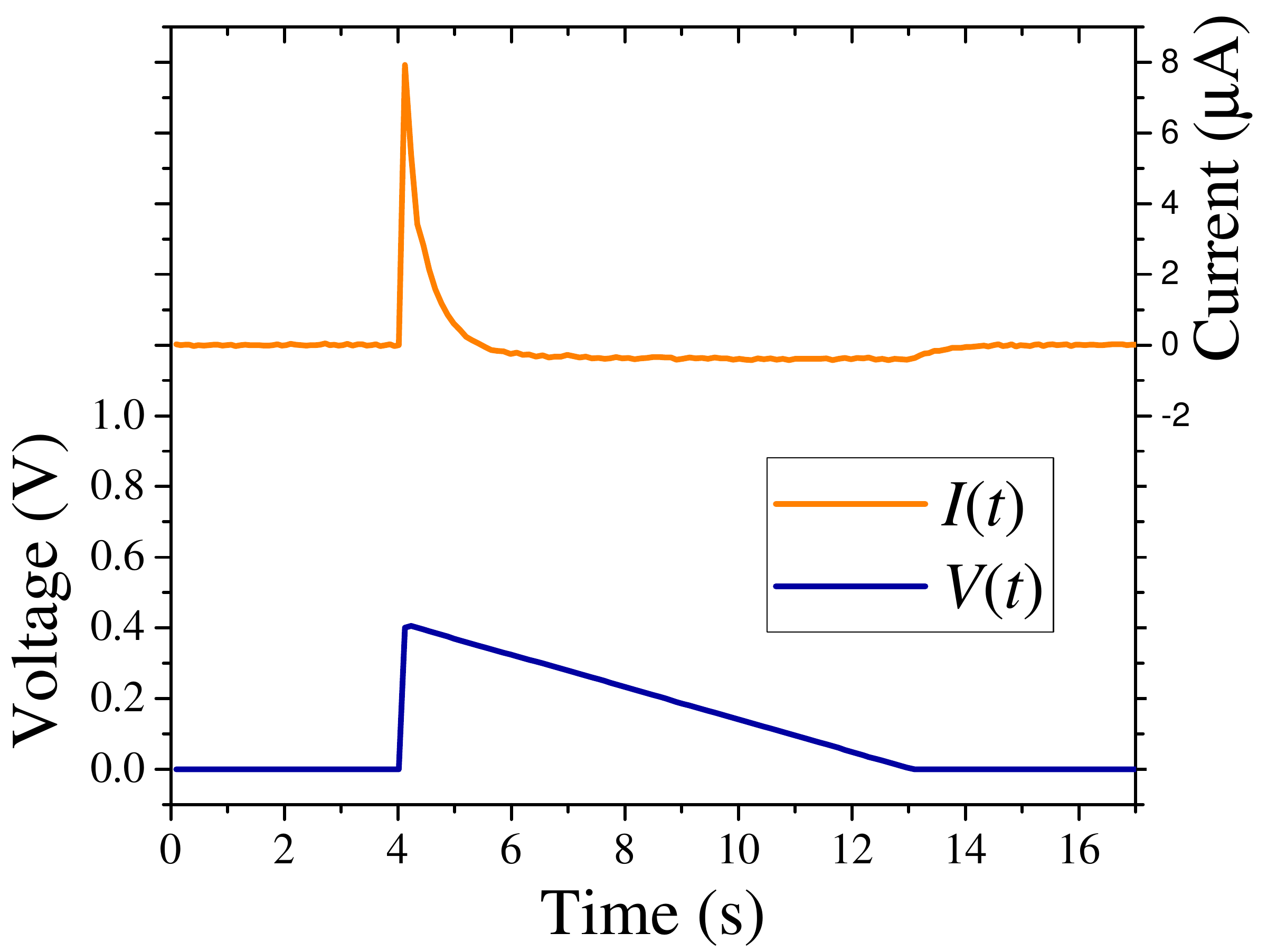} \; \includegraphics[width=28mm]{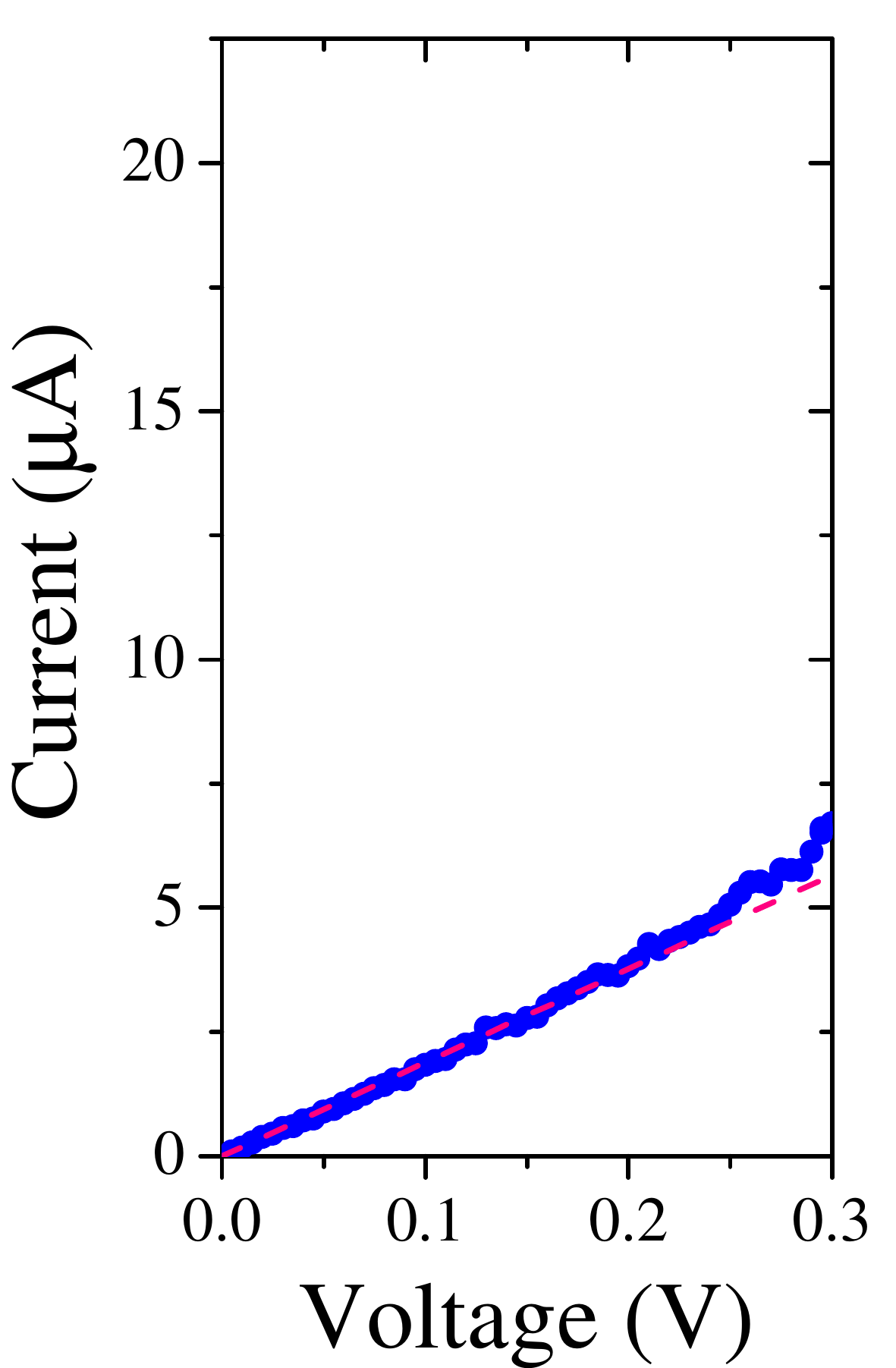}}
(a) \hspace{37mm} (b) \hspace{37mm} (c)
	\caption{Ideal memristor test performed at $V_0=0.4$~V, $C=10$~$\mu$F. (a) A low-amplitude sweep is used to test the initial memristance (the relay is closed). (b) Voltage
		and current versus time, when the testing voltage is applied (the relay is open). (c) A low-amplitude sweep is used to test the final memristance (the relay is closed). The fitting lines in (a) and (c) correspond to the same $R_M=53$~k$\Omega$.
	}
	\label{fig:3}
\end{figure*}

\begin{figure*}[tb]
	\centering{\includegraphics[width=28mm]{fig3c} \includegraphics[width=58mm]{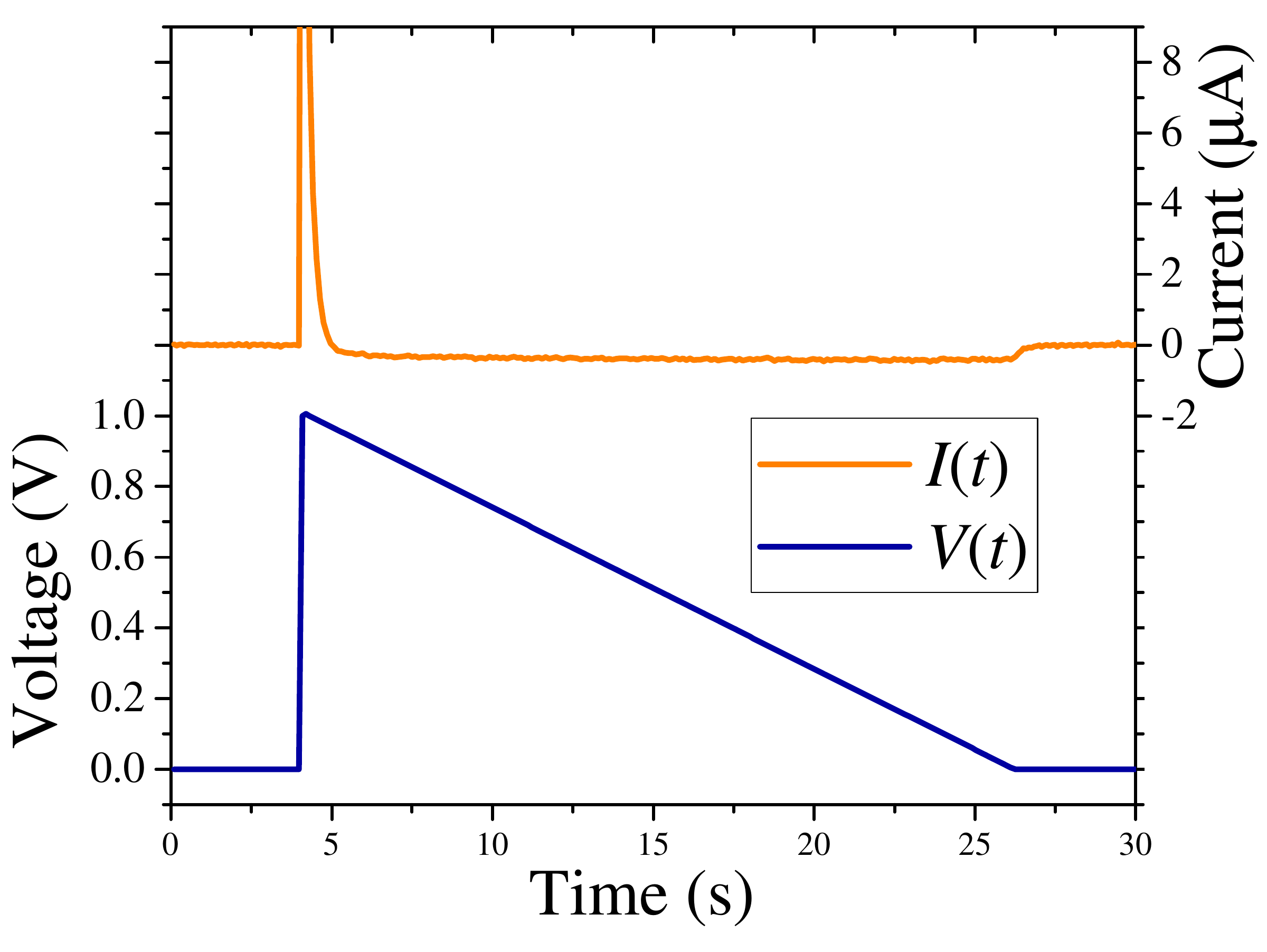}  \includegraphics[width=28mm]{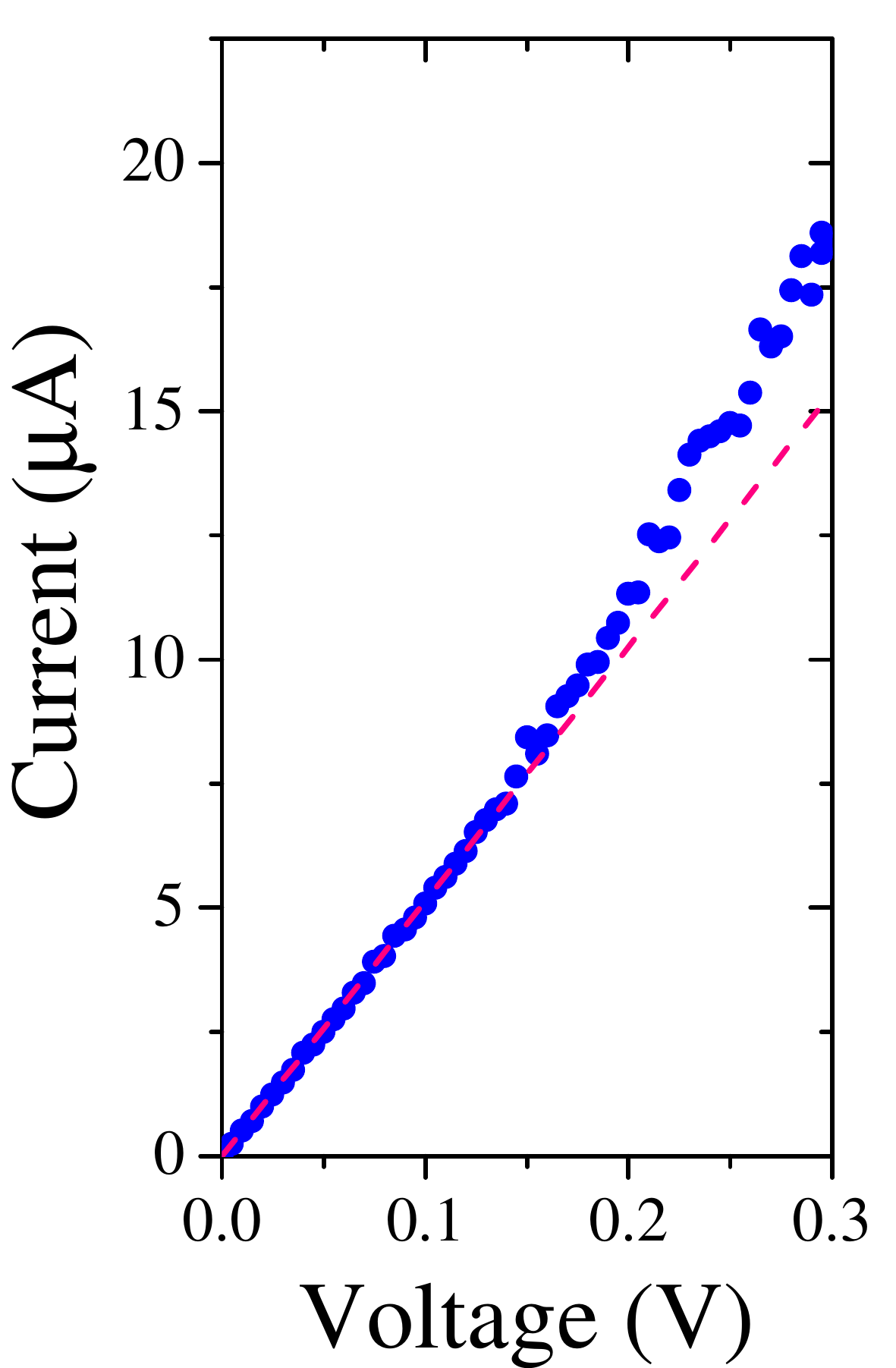}}
(a) \hspace{37mm} (b) \hspace{37mm} (c)
	\caption{Ideal memristor test performed at $V_0=1$~V, $C=10$~$\mu$F. (a) A low-amplitude sweep is used to test the initial memristance (the relay is closed). (b) Voltage
		and current versus time, when the testing voltage is applied (the relay is open). (c) A low-amplitude sweep is used to test the final memristance (the relay is closed). The fitting line in (a) corresponds to $R_M=53$~k$\Omega$, while in (c) to $R_M=19.5$~k$\Omega$.}
	\label{fig:4}
\end{figure*}

The ideal memristor test, as represented in Figure~\ref{fig:1}, was performed at several values of the  pulse amplitude $V_0$ (see Figure~\ref{fig:1}(b)) with the initial memristance set to $R_M=53$~k$\Omega$. Here, we present the results obtained at $V_0=0.4$~V and $V_0=1$~V. Since these measurements were performed in sequence, the final state after the application of $V_0=0.4$~V served as the initial state for $V_0=1$~V. We emphasize that according to the test procedure, the initial and final charge on the capacitor is the same (zero). Therefore, if the tested device were a memristor, its final and initial memristance would be the same too.

It is found that for $V_0=0.4$~V the final memristance is the same as the initial one (cf. Figs. \ref{fig:3}(a) and \ref{fig:3}(c)). However, the larger value of $V_0=1$~V causes the device to switch into the lowest resistance state, $R_{on}=19.5$~k$\Omega$, see Figure~\ref{fig:4}. As the final device state is {\it different} from the initial one when the capacitor has discharged, we conclude that {\it our device has not passed the ideal memristor test}. We observed
similar results for all the other samples tested. Therefore, {\it none} of our devices have passed the ideal memristor test.

We note that
the test was performed using 40~$\mu$A current compliance, which was not exceeded during the test. However, due to the nature of our test, its conclusions are {\it independent} of whether the current was limited or not by the compliance current. Moreover, the transition regions in the $I(t)$ curves at  $t\sim 13.5$~s in Fig. \ref{fig:3} and $t\sim 26$~s in Fig. \ref{fig:4} correspond to the capacitor discharge process. Zero current at the final moment of time indicates that the capacitor has been discharged.

\subsubsection{Commercially available devices}

\begin{figure*}[t]
    \centering{(a)\includegraphics[width=90mm]{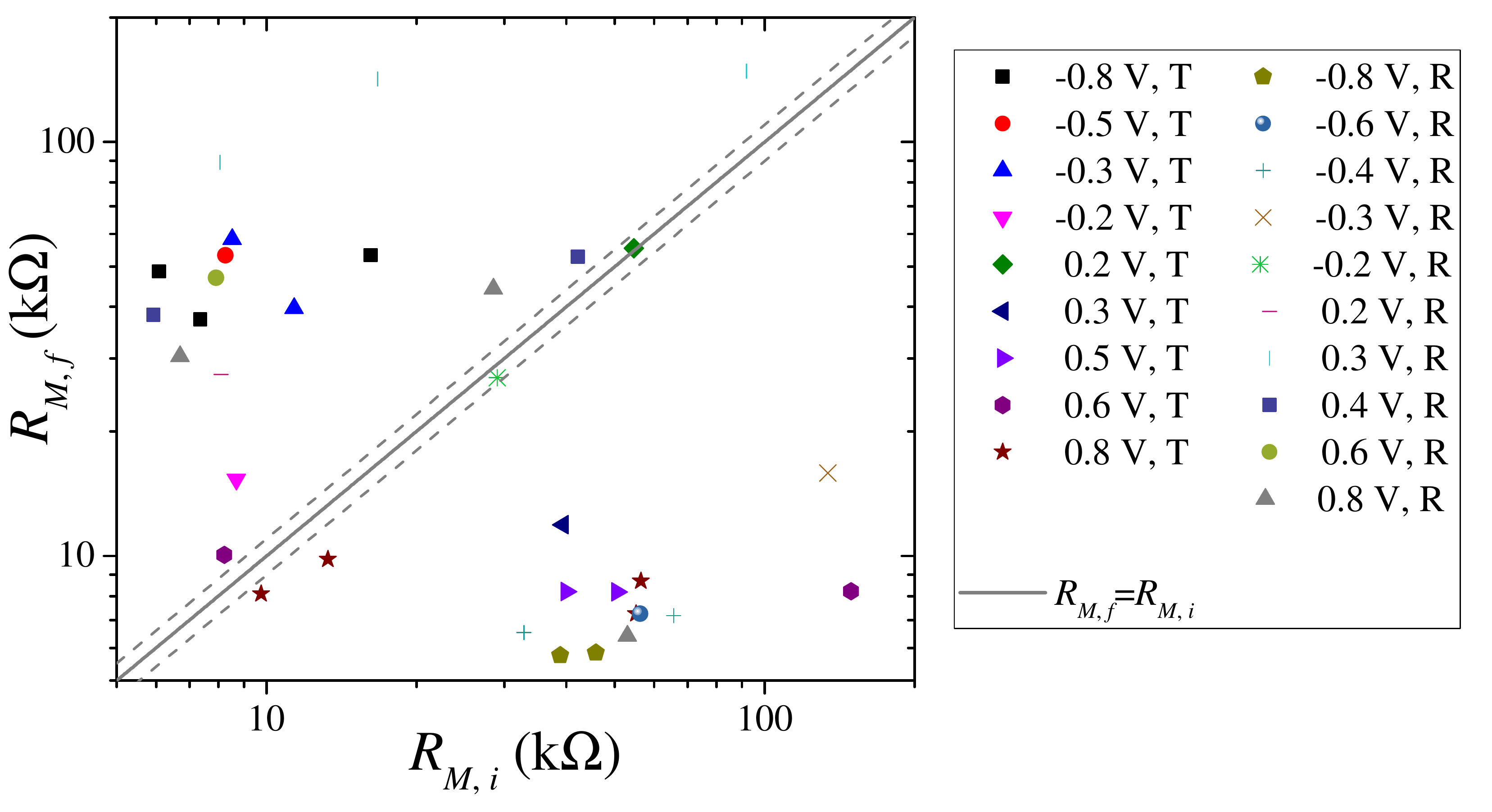}}
    \centering{(b)\includegraphics[width=90mm]{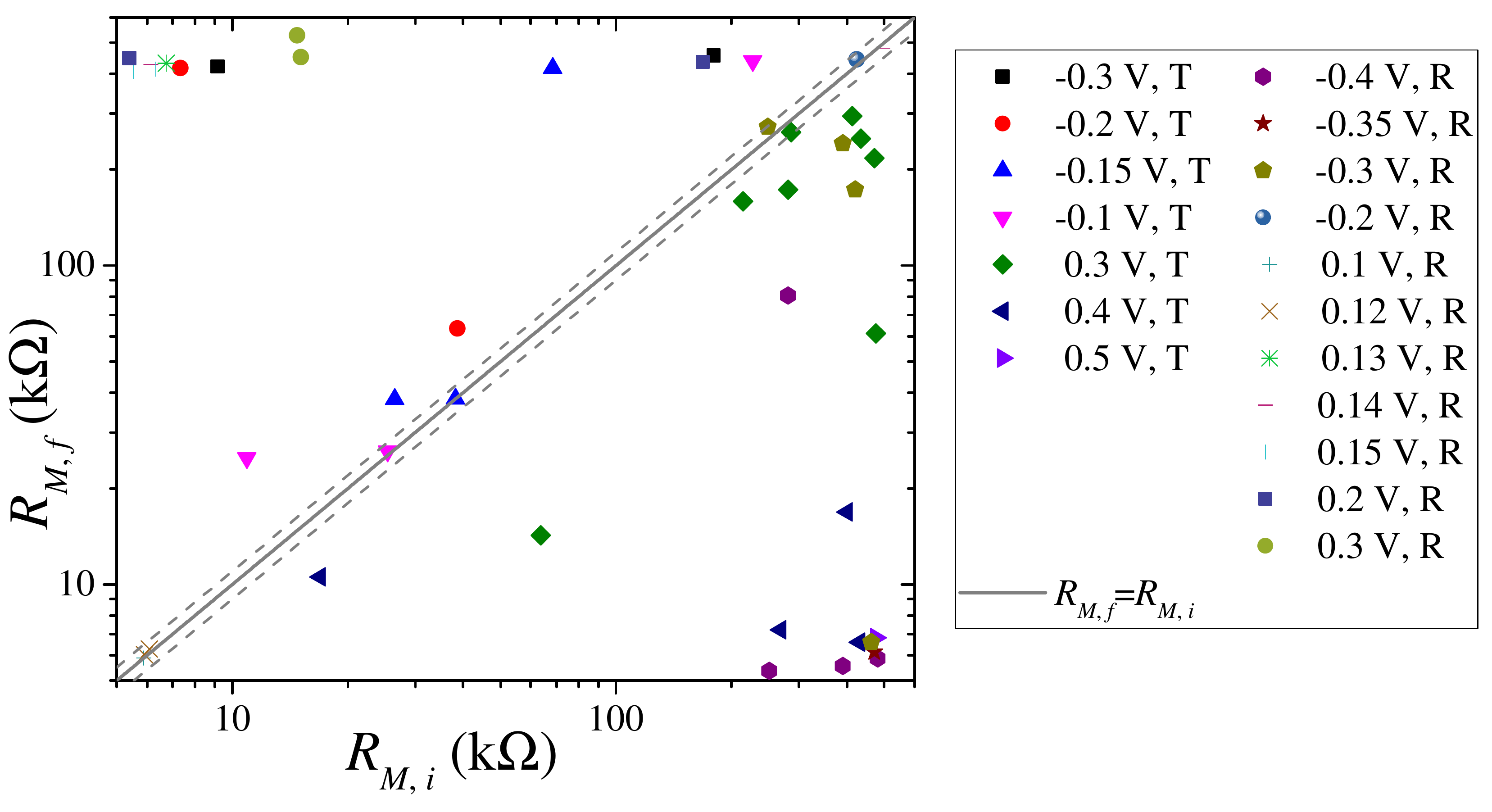}}
	\caption{ Final versus initial resistance found in a series of tests (with $C=1$~$\mu$F) using triangular (T, as in Fig.~\ref{fig:1}) and rectangular (R) pulses of positive and negative polarities. Results for BS-AF-W and M+SDC~Cr  devices are shown in (b) and (c), respectively. The dashed lines correspond to 10~\% deviation from $R_{M,f}=R_{M,i}$ line.}
	\label{fig:5}
\end{figure*}

We have also applied the memristor test to commercially available EMCs~\cite{knowm} (Knowm, Inc,).  Their operation is based on the movement of Ag atoms through a stack of chalcogenide layers with one of Ge$_2$Se$_3$ layers doped either by W (BS-AF-W devices) or Cr (M+SDC~Cr devices). The current-voltage curves of the samples used in the memristor test are presented in Fig.~\ref{fig:2}(b) and (c). Their form indicates the bipolar resistance switching mode, similar to the one in Fig.~\ref{fig:2}(a). We note that the switching thresholds of the Knowm devices are smaller than those in our Cu-SiO$_2$ devices. Moreover, a large $R_{off}\sim 500$~k$\Omega$ was observed in the M+SDC~Cr device. To reduce the capacitor discharge time, we used a smaller $1$~$\mu$F capacitor in experiments with the Knowm devices.

A series of more than 30 tests were applied to each BS-AF-W and M+SDC~Cr sample. In these tests, we used the triangular (like in Fig.~\ref{fig:1}(b)) and rectangular pulses of positive and negative voltage. The width of restangular pulses was 2~s, the slope of triangular pulses was about 0.025~s. The resistance was measured using a 10~mV voltage, the waiting time before the final resistance measurement was $\geq$~3~s. The pulses were applied in an arbitrary order, and the results of these measurements are summarized in Fig.~\ref{fig:5}.

Each point in Fig.~\ref{fig:5} corresponds to a single measurement like the one in Fig.~\ref{fig:3} or \ref{fig:4}. For  reference, the straight line represent the condition of equal initial and final states, $R_{M,f}=R_{M,i}$, and dashed lines correspond to 10~\% deviations from this condition. So, if the device under test were a memristor, the measurement results would group between the dashed lines.  Clearly, it is not the case of {\it both} BS-AF-W and
M+SDC~Cr electrochemical metallization cells. In agreement with the results of Fig.~\ref{fig:4}, the tendency of $R_{M,f}<R_{M,i}$ for positive triangular pulses, and $R_{M,f}>R_{M,i}$  for negative triangular pulses can be recognized. In the case of rectangular pulses the tendency is opposite as the final state is significantly defined by the falling front of the pulse. Overall, as most of the data point are outside of the dashed line interval, the conclusion is that also the Knowm devices are not memristors.

\subsection{Valence change memory cells}

\begin{figure*}[t]
    \centering{(a)\hspace{-5mm}\includegraphics[width=59mm]{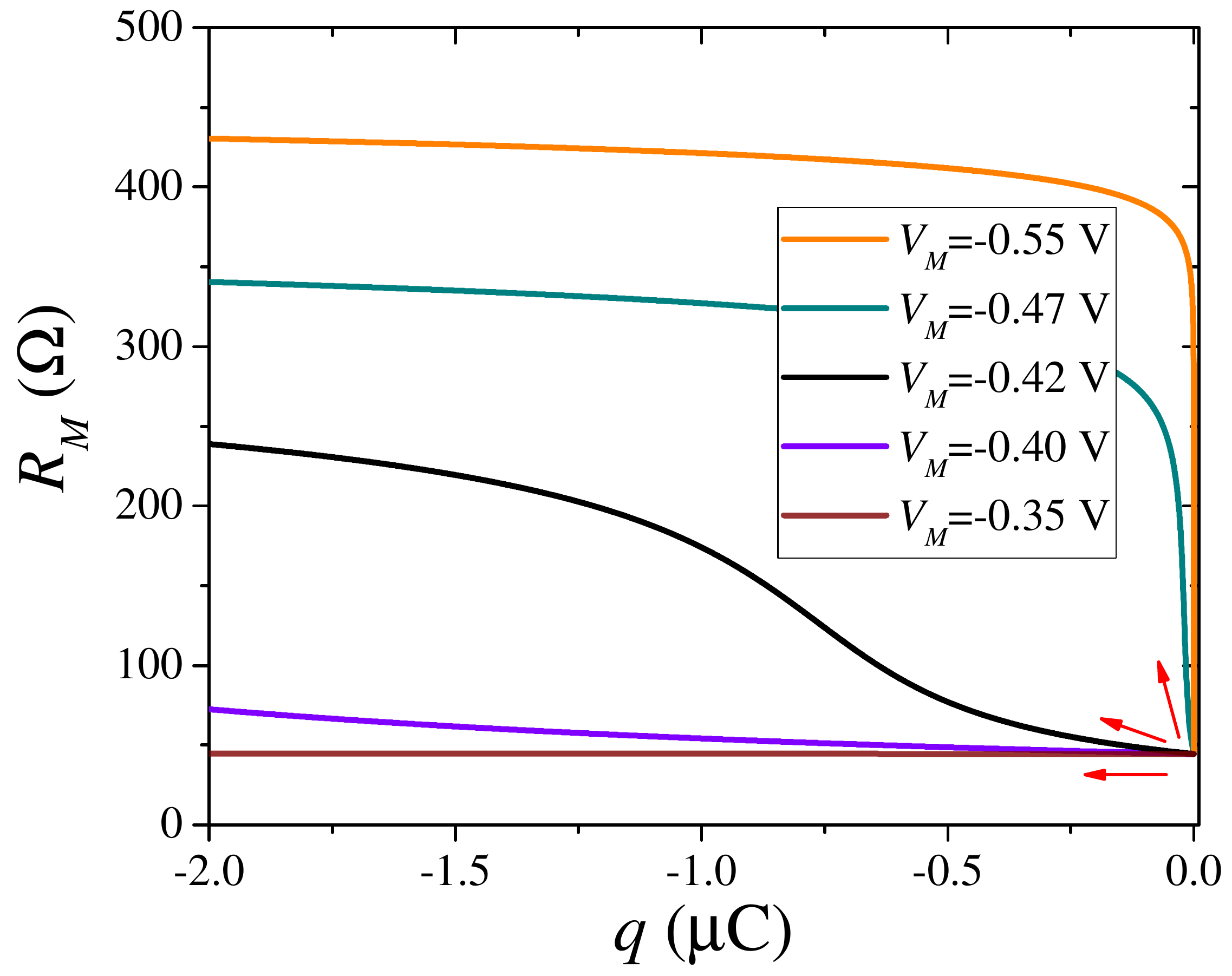} (b)\hspace{-5mm}\includegraphics[width=59mm]{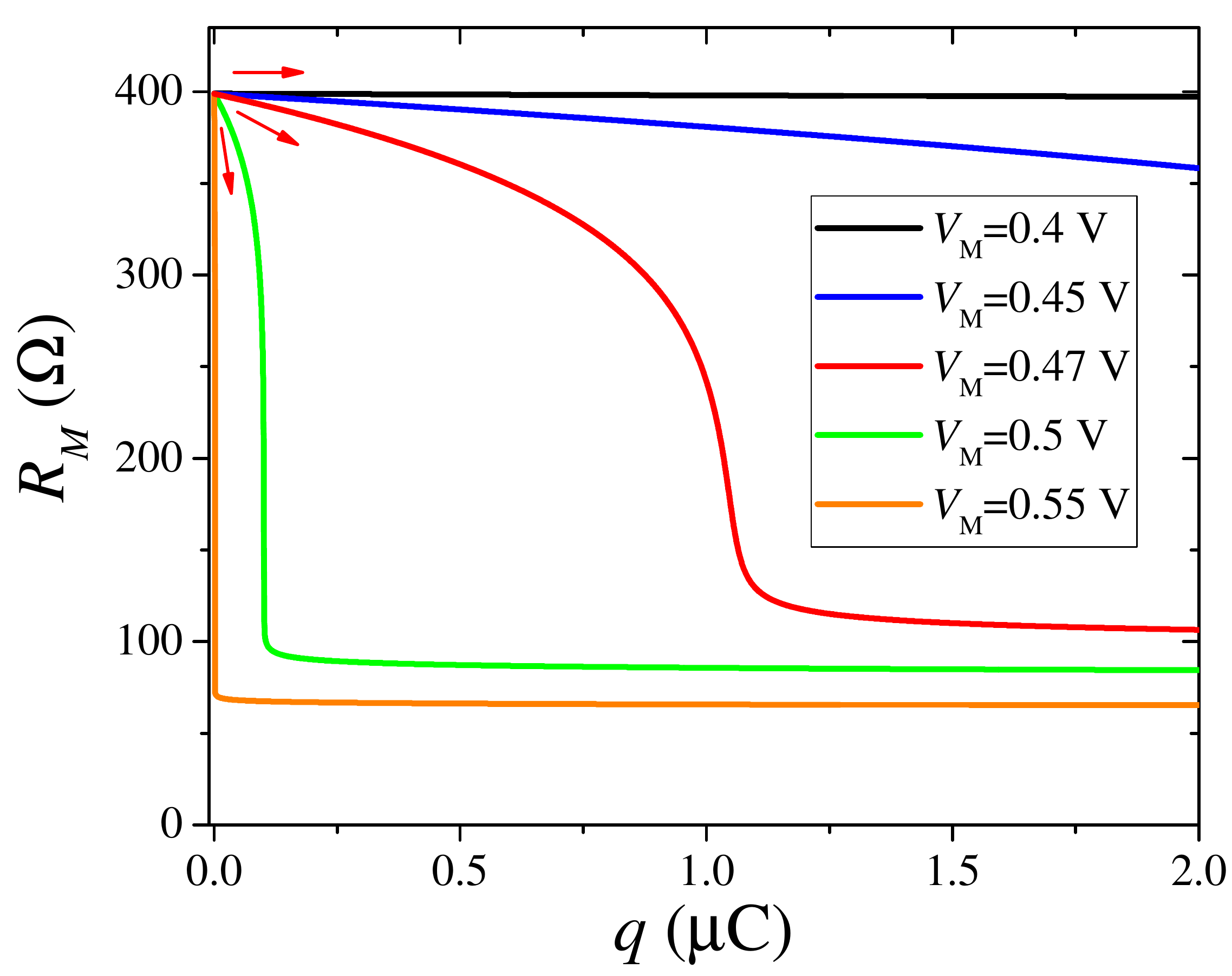}}
	\caption{ Resistance as function of charge that flows through a TaO device found for the cases of (a) negative and (b) positive  constant
voltages. These plots were obtained using Eqs. (\ref{eq:1})-(\ref{eq:2}) model with $x(t=0)=0.9$ in (a) and $x(t=0)=0.1$ in (b).
}
	\label{fig:6}
\end{figure*}

As VCMs are the second major class of resistive memory devices, it is of interest to understand how close their behavior is to the memristor model. For this purpose, we consider the transient dynamics of TaO cells, a representative of VCMs devices, and employ Eqs. (\ref{eq:1})-(\ref{eq:2}) to discuss their dynamics, as these have been shown to accurately reflect the experimental data~\cite{Strachan13a}. Figure~\ref{fig:6} shows the resistance as a function of charge that flows through the cell found for a set of applied voltages. One can notice that starting from the same point, the $R_M(q)$ curve actually splits into individual curves for each voltage. Clearly, there is no single $R_M(q)$ that ideal memristors would satisfy, or a grouping around a certain $R_M(q)$ that could be taken into account by small corrections to the ideal  model.

\section{Discussion}
\subsection{Resistance-switching memories are definitely not memristors}

We can further expand on these experimental results as follows. In this work we have applied the ideal memristor test suggested in Ref.~\cite{pershin18a} to in-house fabricated Cu-SiO$_2$-based ECMs and commercially available Ag-based ECMs, which are a type of resistance-switching memories. As part of the test, we have compared the initial device states with the final ones obtained under the condition of a
capacitor discharge in series with the memory device. Since in multiple tests and for a wide range of driving conditions 
 we have found that the final states of the memory devices were quite different from the initial ones, we conclude that these resistance-switching devices cannot be described simply by a memristance that depends on the charge only: $R_M(q)$. Therefore, they are {\it not} memristors. Since the current-voltage characteristics of the devices used in our study are typical of a wide range of resistance-switching cells, our general conclusion is that {\it resistance-switching memories are not memristors}, irrespective of their specific device structures and switching mechanisms.

We also note that the triangular-shape voltage signal $V(t)$ employed in our work has facilitated the ideal memristor test. Under the test conditions of Figure~\ref{fig:1}, the tested devices were subjected first to a relatively large positive voltage (the initial magnitude is $V_0$), followed by a small negative voltage. The tested devices failed the test since the positive voltage across the devices was sufficient to switch $R_M$ to $R_{on}$, while the negative voltage was not sufficient for the inverse switching. Since, in principle, for any given pair of $V_{t,+}$ (positive voltage) and $V_{t,-}$ (negative voltage), one can always choose the test signal such that $V_0>V_{t,+}$ and $V_M(t)>V_{t,-}$, one can further argue that {\it there are no ideal memristors among the threshold-type resistance-switching memory cells}.

\subsection{Memristor impossibility conjectures}

At this juncture, the reader may ask how the ideal memristor model is related, if at all, to physical resistance-switching devices such as those studied experimentally in this work, or any other similar devices published in the literature (see, e.g., references in~\cite{pershin11a}). Accounting for the fact that the response of physical devices is different from the ideal behavior, Chua argued~\cite{chua2011a} that the resistance-switching memories are an ``unfolding'' theory extension of the ideal devices. More recently, he also proposed~\cite{Chua_2019a} that the resistance-switching can be represented by a circuit combining ideal memristors with some other ideal devices~\cite{chua03a}.
However, these statements are clearly {\it incorrect}.

In fact, unfolding relies on families of mathematical functions that are similar (close) to
each other. When an idealized model is partially inadequate, the model can be improved by adding {\it small corrections} resulting in the new model: an unfolding of the original system~\cite{Murdock:2006,bruce_giblin_1992}. However, this approach is {\it not} applicable to the physical (experimentally-realizable) memristive devices because the difference between their physical models (as known in the literature) and the ideal memristor model can not be bridged by small correction terms. Similarly, a circuit representation of physical memory devices by circuits of ideal components is highly unlikely for the same reason: the ideal memristor behavior is too different from that of physical devices.

Based on the above arguments, we formulate two {\it memristor impossibility conjectures} that may serve as foundations for future research.

\begin{itemize}
  \item[] {\bf First memristor impossibility conjecture.} It is impossible to accurately model physical resistance-switching memories by adding small corrections to the {\it ideal} memristor model.
  \item[] {\bf Second memristor impossibility conjecture.} It is impossible to  accurately model physical resistance-switching memories by a circuit  combining {\it ideal} memristors with any kinds of non-linear {\it ideal} circuit elements.
\end{itemize}

In the second conjecture, we refer to the ideal elements defined in Ref.~\cite{chua03a}. It can be also formulated in the strong sense considering only the combinations of memristors with basic circuit elements (non-linear resistors, capacitors, and inductors).

\section{Conclusions}
In conclusion, we have employed a recently suggested test~\cite{pershin18a} to experimentally verify whether currently existing resistance-switching memories are indeed
memristors, as it was claimed in Ref.~\cite{chua2011a}, or not. Our results demonstrate {\it unambiguously} that {\it they are not}.

Unlike the behavior of ideal memristors,
the final states of the memory devices we have measured {\it significantly} deviate from their initial states. These 
deviations {\it cannot} be accounted for by small corrections to the ideal memristor relations. 
 
This study has then led us to formulate two conjectures on the impossibility of building
a model of physical (experimentally-realizable) resistance-switching memories based on the ideal memristor behavior. The collection of these
experimental results cast further doubts on the existence of the ideal memristor as a forth circuit element that can be fabricated experimentally.
In fact, various previous results reported in the literature such as, e.g., threshold-type hysteresis curves of physical devices, stochastic
switching~\cite{gaba2014memristive,Naous16a}, and CRS behavior~\cite{linn2010complementary,6850077} are not compatible with the memristor model and thus support our conclusions.

\section*{Acknowledgment}
The authors are thankful to Mr. H.~Smith and Dr. T.~M.~Crawford for their help with sputtering deposition. This work was partially supported by a USC Provost grant.

\bibliographystyle{advancedmaterials_YP}
\bibliography{memcapacitor}

\begin{thebibliography}{10}
\providecommand{\url}[1]{\texttt{#1}}
\providecommand{\urlprefix}{URL }

\bibitem{strukov08a}
D.~B. Strukov, G.~S. Snider, D.~R. Stewart, R.~S. Williams, \emph{Nature}
  \textbf{2008}, \emph{453}, 80.

\bibitem{Wang19a}
F.~Z. Wang, L.~Li, L.~Shi, H.~Wu, L.~O. Chua, \emph{Journal of Applied Physics}
  \textbf{2019}, \emph{125}, 054504.

\bibitem{chua71a}
L.~O. Chua, \emph{{IEEE} Trans. Circuit Theory} \textbf{1971}, \emph{18}, 507.

\bibitem{chua2011a}
L.~Chua, \emph{Applied Physics A} \textbf{2011}, \emph{102}, 765.

\bibitem{mouttet2012memresistors}
B.~Mouttet, \emph{arXiv preprint arXiv:1201.2626} \textbf{2012}.

\bibitem{meuffels2012fundamental}
P.~Meuffels, R.~Soni, \emph{arXiv preprint arXiv:1207.7319} \textbf{2012}.

\bibitem{di2013physical}
M.~Di~Ventra, Y.~V. Pershin, \emph{Nanotechnology} \textbf{2013}, \emph{24},
  255201.

\bibitem{vongehr2015missing}
S.~Vongehr, X.~Meng, \emph{Scientific Reports} \textbf{2015}, \emph{5}, 11657.

\bibitem{sundqvist2017memristor}
K.~M. Sundqvist, D.~K. Ferry, L.~B. Kish, \emph{Fluctuation and Noise Letters}
  \textbf{2017}, \emph{16}, 1771001.

\bibitem{comment_19a}
Y.~V. Pershin, M.~{Di Ventra}, \emph{Semiconductor Science and Technology}
  \textbf{2019}, \emph{34}, 098001.

\bibitem{chua2014if}
L.~Chua, \emph{Semiconductor Science and Technology} \textbf{2014}, \emph{29},
  104001.

\bibitem{chua76a}
L.~O. Chua, S.~M. Kang, \emph{Proc. {IEEE}} \textbf{1976}, \emph{64}, 209.

\bibitem{pershin18a}
Y.~V. Pershin, M.~{Di Ventra}, \emph{J. Phys. D: Appl. Phys.} \textbf{2018},
  \emph{52}, 01LT01.

\bibitem{valov2011electrochemical}
I.~Valov, R.~Waser, J.~R. Jameson, M.~N. Kozicki, \emph{Nanotechnology}
  \textbf{2011}, \emph{22}, 254003.

\bibitem{pershin11a}
Y.~V. Pershin, M.~Di~Ventra, \emph{Advances in Physics} \textbf{2011},
  \emph{60}, 145.

\bibitem{waser2009redox}
R.~Waser, R.~Dittmann, G.~Staikov, K.~Szot, \emph{Advanced materials}
  \textbf{2009}, \emph{21}, 2632.

\bibitem{Strachan13a}
J.~P. Strachan, A.~C. Torrezan, F.~Miao, M.~D. Pickett, J.~J. Yang, W.~Yi,
  G.~Medeiros-Ribeiro, R.~S. Williams, \emph{IEEE Transactions on Electron
  Devices} \textbf{2013}, \emph{60}, 2194.

\bibitem{knowm}
Self directed channel memristors.
\newblock \url{https://knowm.org/downloads/Knowm{\_}Memristors.pdf}.
\newblock Accessed: 2020-03-20.

\bibitem{schindler2007bipolar}
C.~Schindler, S.~C.~P. Thermadam, R.~Waser, M.~N. Kozicki, \emph{IEEE
  Transactions on Electron Devices} \textbf{2007}, \emph{54}, 2762.

\bibitem{Ascoli17a}
A.~Ascoli, V.~Ntinas, R.~Tetzlaff, G.~C. Sirakoulis, \emph{Electronics Letters}
  \textbf{2017}, \emph{53}, 1125.

\bibitem{Chua_2019a}
L.~O. Chua, \emph{Semiconductor Science and Technology} \textbf{2019},
  \emph{34}, 098002.

\bibitem{chua03a}
L.~O. Chua, \emph{Proc. {IEEE}} \textbf{2003}, \emph{91}, 1830.

\bibitem{Murdock:2006}
J.~Murdock, \emph{Scholarpedia} \textbf{2006}, \emph{1}, 1904, revision
  \#91898.

\bibitem{bruce_giblin_1992}
J.~W. Bruce, P.~J. Giblin.
\newblock \emph{Curves and Singularities: {A} Geometrical Introduction to
  Singularity Theory}.
\newblock Cambridge University Press, 2 edition, \textbf{1992}.

\bibitem{gaba2014memristive}
S.~Gaba, P.~Knag, Z.~Zhang, W.~Lu.
\newblock In \emph{2014 IEEE International Symposium on Circuits and Systems
  (ISCAS)}. IEEE, \textbf{2014} 2592--2595.

\bibitem{Naous16a}
R.~{Naous}, M.~{Al-Shedivat}, K.~N. {Salama}, \emph{IEEE Transactions on
  Nanotechnology} \textbf{2016}, \emph{15}, 15.

\bibitem{linn2010complementary}
E.~Linn, R.~Rosezin, C.~K{\"u}geler, R.~Waser, \emph{Nature materials}
  \textbf{2010}, \emph{9}, 403.

\bibitem{6850077}
E.~{Linn}, A.~{Siemon}, R.~{Waser}, S.~{Menzel}, \emph{IEEE Transactions on
  Circuits and Systems I: Regular Papers} \textbf{2014}, \emph{61}, 2402.

\end{thebibliography}
\end{document}